\documentclass{article}
\usepackage{graphicx} 
\usepackage[a4paper, total={6in, 8in}]{geometry}
\date{September 2025}

\usepackage{xcolor}
\usepackage[version=3]{mhchem}
\begin{document}


\title{Bound-state-in-the-continuum (BIC) induced narrow resonances in MXene-coated absorptive dielectric metasurfaces for Methane sensing}

\maketitle


\author{Shubhanshi Sharma$^1$*},
\author{Monica Pradhan$^2$}
\author{Aviad Katiyi$^3$}, 
\author{Alina, Karabchevsky$^{3,4}$},
\author{Shailendra Kumar Varshney$^1$}


Email Address: $^*$shubhanshi07@iitkgp.ac.in\\
$^1$Department of Electronics and Electrical Communication Engineering, Indian Institute of Technology Kharagpur, Kharagpur, West Bengal, India, 721302\\
$^2$School of Nanoscience and Technology, IIT Kharagpur, West Bengal, 721302, India \\
$^3$School of Electrical and Computer Engineering, Ben-Gurion University of the Negev, Beer-Sheva, 8410501, Israel\\
$^4$Department of Physics, Lancaster University, LA1 4YB, United Kingdom


\textbf{Keywords}: Bound States in the continuum, MXene, Hybrid Metasurfaces, Fano resonances

\begin{abstract}
Strong light confinement is highly necessary for various applications, including sensing. MXene, a novel and emerging material with a broadband plasmonic response, has been highly utilized in electronic sensing systems, as well has garnered significant attention for its applicability in photonics. The loss imparted by MXene can be overcome through the Bound states in Continuum (BIC) physics. In this work, we report two important designs of a hybrid metasurface, comprising a silicon nanodisk metasurface and MXene. In both designs, narrow and high absorption resonance of quality factor of $\sim$150 is attained, where MD is the dominant multipole, governed by symmetry-protected BIC. Both hybrid metasurface designs are optimized to exhibit narrow resonance in the vicinity of 1650 nm with an absorption greater than 90 $\%$. The origin of high absorbance in such a hybrid metasurface is attributed to the momentum matching by the spacer layer of $\ce{SiO_2}$. The spectral characteristics of the designed metasurface can be utilized for first overtone spectroscopy of Methane gas. Numerical simulations yield a bulk refractive index sensitivity of 171 nm/RIU with FOM = 17.56 \ce{RIU^{-1}} and sensitivity for Methane gas, S = 0.8 nm per unit percentage concentration, when a Cryptophane-E layer is used.

\end{abstract}

\section{Introduction}
Over the past decades, artificially engineered structures named metamaterials have gained attention due to their unique abilities to manipulate electromagnetic wave amplitude and phase fronts \cite{karabchevsky2020chip}. The two-dimensional variants are referred to as metasurfaces, where individual elements (meta-atoms) couple resonantly with electric or magnetic or both components of the incident field and exhibit a response not found in nature \cite{hsiao2017fundamentals}. With these versatile properties, metasurfaces demonstrate advantages in various applications such as meta-lensing \cite{khorasaninejad2017metalenses}, cloaking \cite{chen2012invisibility}, holography \cite{jiang2019metasurface}, sensing \cite{qin2022metasurface}, beam steering \cite{lin2022high}, high-harmonic generation \cite{sain2019nonlinear}, absorbers \cite{badloe2017metasurfaces}, etc.  
\begin{figure*}[ht!]
 \centering
 \includegraphics[width=\linewidth]{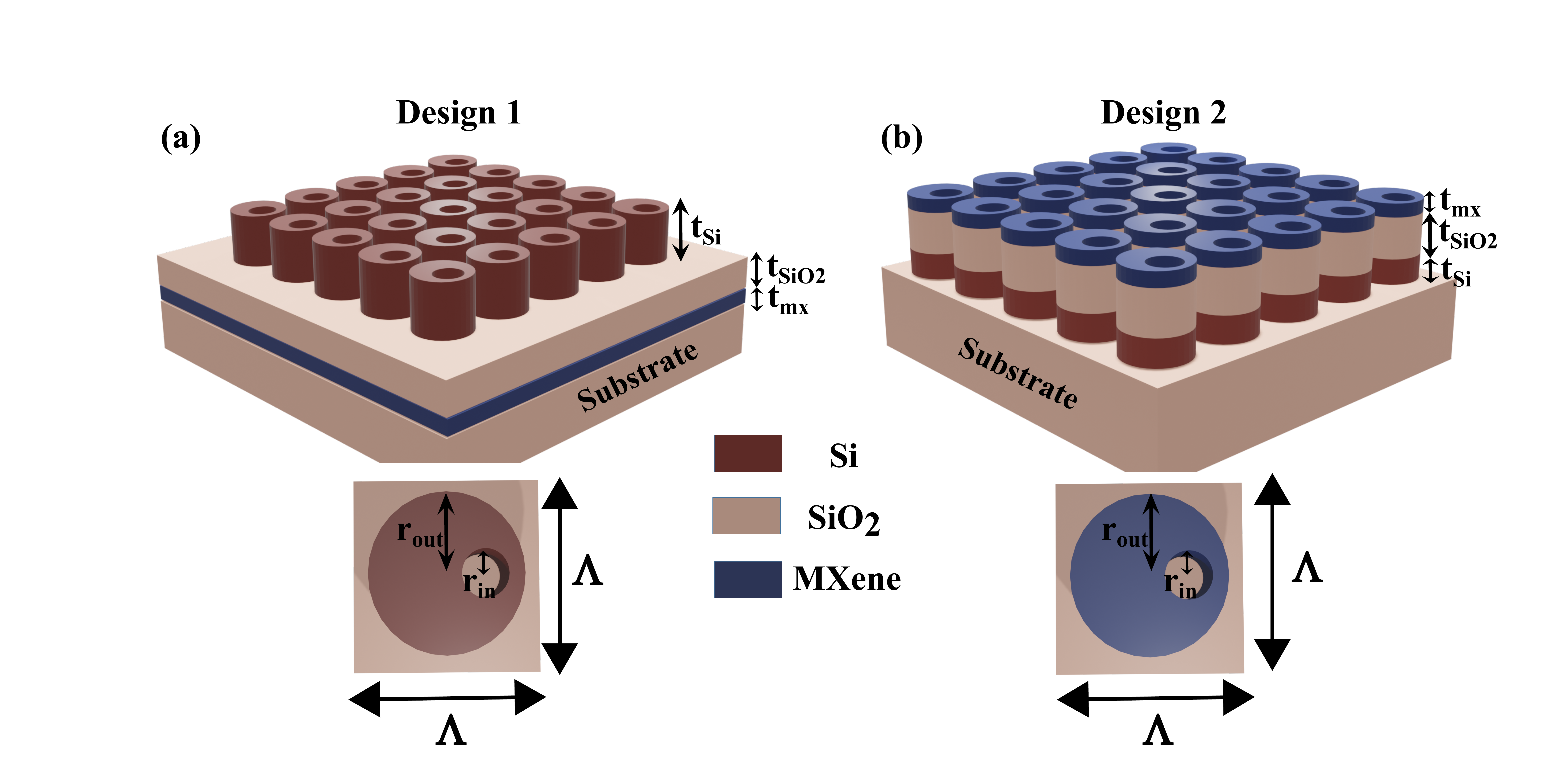}
 \caption{3D Schematics of hybrid metasurface absorber with MXene in two design configurations, while the top view of a meta-atom is also given below corresponding to both designs, (a) Design 1:  MXene is used as an absorptive plasmonic layer sandwiched between the Silica as a spacer layer and substrate with the Si nanodisk. The silicon nanoresonator has radius ($r_{out}$) = 268 nm and comprises a thorough air channel of radius ($r_{in}$) = 160 nm. Other design parameters are, thickness ($t_{Si}$) = 242 nm, periodicity, $\Lambda$ = 865 nm, thickness of $\ce{SiO_2}$ spacer layer ($t_{SiO_2}$) is 200 nm, and the thickness of the MXene layer ($t_{mx}$) is 90 nm. (b) Design 2: MXene is used as a nanoresonator. The design configuration consists of a composite nandisk resonator, made from Mxene, $\ce{SiO_2}$, and Si, placed on a silica substrate. The composite nanodisk also has an air channel similar to design 1. The silica space layer thickness, $t_{SiO_2}$ = 320 nm, while other parameters are similar to design 1. In both the design configurations, the air channel offset in x direction is quantified by $\Delta s$}
  \label{fig1}
\end{figure*}

 
 \begin{figure*}[ht!]
 \centering
 \includegraphics[width=\linewidth]{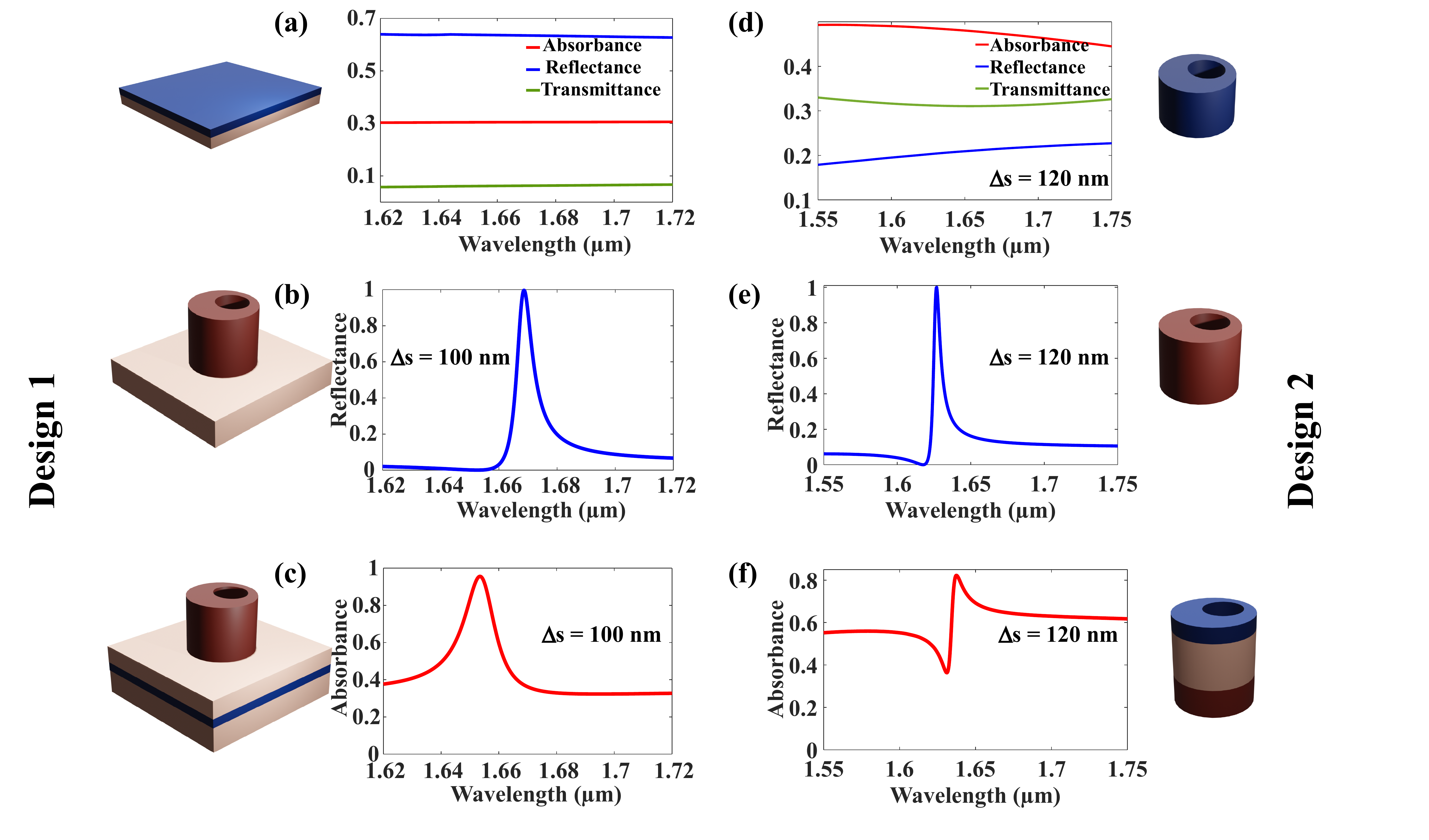}
 \caption{Comparison between the optical responses of both designs. Design 1: Optical response of (a) a thin layer of MXene, (b) only a Si nanodisk, and (c) a hybrid nanostucture over the Silica substrate. Design 2: Optical response of (a) MXene nanodisk, (b) Si nanodisk, and (c) hybrid nanodisk without substrate.  }
 \label{fig2}
\end{figure*}
Highly efficient and nearly perfect absorbers are widely applied in sensing, energy harvesting \cite{wang2015highly}, scattering reduction, modulation \cite{cheng2023large}, IR camouflage, and wireless communication. In the case of optical sensing, light absorption is desirable to a greater extent, as it suppresses reflection and transmission, allowing energy to dissipate in the absorbing layer. The localized surface plasmon resonance (LSPR) \cite{liu2010infrared} has been employed to create a resonant or a broadband absorber in plasmonic metasurfaces \cite{watts2012metamaterial,li2014refractory}. Alternatively, an all-dielectric metasurface supports two Mie resonances (electric and magnetic) whose overlap results in super and suppressed optical absorption \cite{tian2020high}. 
Therefore, it is possible to achieve absorption of both electric and magnetic fields by suitable design and proper material selection \cite{tao2008metamaterial}. The interference between multipoles results in non-radiating states \cite{pradhan2023engineered,sharma2023visible}, one of which is the Bound state in the continuum (BICs) that possesses an indefinite lifetime. The BICs are the states confined within the continuum and remain decoupled from it \cite{hsu2016bound}\cite{PhysRevB.78.075105}. These states ideally possess an infinite quality (Q) factor and cannot be observed in transmittance or reflectance spectra. BICs are of two types: one is governed by the system's symmetry, known as symmetry-protected BIC (SPBIC) \cite{PhysRevLett.107.183901}, and the other arises due to the destructive interference of the states, known as accidental BICs \cite{PhysRevLett.100.183902}\cite{PhysRevLett.113.037401}. Perturbations in geometry are generally introduced to observe the BICs in the spectra, transforming the non-radiative BICs into radiative quasi-BICs with a finite but high Q-factor exhibiting either Fano resonance or electromagnetically induced transparency (EIT) \cite{https://doi.org/10.1002/adom.201900383}\cite{Sharma_2023}. Each resonance can absorb the incident energy by achieving a critical coupling state, where the radiative loss rate is the same as the material loss \cite{doi:10.1021/acsphotonics.2c00901}. Hybrid multilayer structure-based broadband absorbers are crafted in a way that works in visible \cite{zhao2020lithography} and mid-infrared \cite{yu2019broadband} wavelength ranges. Besides these, metasurface absorbers have also been explored in the Gigahertz \cite{guo2024double}, Terahertz \cite{liu2023plasmonic}, and Ultraviolet \cite{abou2023ito} regions. \\
 The hybrid structure consists of metal and dielectric, which can yield near-unity absorption. Within this context, metals such as gold, silver, copper, and tungsten are widely used to make tunable devices that operate in different frequency bands \cite{rifat2018hybrid}. Such hybrid structures can generate surface plasmon resonance (SPR) and localized surface plasmon resonance (LSPR) due to the oscillation of electrons. The plasmons effectively manipulate light at the sub-diffraction scale, making them valuable for controlling light-matter interaction. 
 Despite promising capabilities, several limitations are associated with metal usage in such devices. These include oxidation, melting, significant intrinsic losses, and high preparation costs. That has led to exploring alternative two-dimensional transition materials, such as MXene. \\
 MXenes belong to the family of metal carbides, nitrides, and carbo-nitrides represented by the general formula $M_{n+1}X_{n}T_{x}$, where M refers to an early transition metal, X is either carbon or nitrogen, and T stands for surface functional groups, usually O, OH, or F \cite{zong2024graphene}. These materials were initially discovered in 2011 at Drexel University by selectively removing the A layer (A-group mostly the elements of group 13 and 14) \cite{gogotsi2023MXenes} from bulk ternary transition metal carbides and nitrides, known as MAX phases, resulting in the formation of multilayered MXenes \cite{oliveira2023structure}. MXenes show high metallic conductivity due to the free electrons in their transition metal carbide or nitride backbone that contribute to their unique electronic structure \cite{zhang2024advanced}. The hydrophilic nature of MXene permits easy deposition on a wide range of substrates. Its anisotropic behavior leads to optical absorption covering broad bandwidths from visible to mid-infrared regions. MXene, as a 2D material, can be modified as photosensitive for the applications of flexible photodetectors \cite{hu2023strategy}. The abundance of functional groups induces dipole and interfacial polarization, contributing to polarization losses. MXene possesses a direct band gap, which can be altered by changing the number of layers or modifying the surface functional groups \cite{khazaei2013novel}. The surface functional groups of MXenes facilitate the absorption, while the large surface area enhances the sensor sensitivity \cite{phuong2023application,zhu2022properties}. Based on these features, optical fiber-based sensors using MXene have gained widespread attention and are extensively utilized in various research fields \cite{chen2020refractive,singh2024waveflex,li2024MXene,pacheco2022MXene}. The investigation of MXenes metasurface for realizing a broadband absorber has been demonstrated \cite{jia2022visible,huang2024inkjet,chaudhuri2018highly}. \\
 MXene exhibits properties that lie between those of true two-dimensional materials and bulk three-dimensional (3D) materials, thereby classifying it as a quasi-2D material \cite{mehmood2023MXene}. Since MXene exhibits plasmonic response in the near infrared region, they are considered an excellent candidate for efficient light absorption due to their stability and tunable surface chemistry \cite{saini2021emerging}. A broadband MXene absorber composed of nanodisk, nanobars, and a multilayer of $\ce{MXene/SiO_{2}}$ has already been explored, showing a large optical absorption bandwidth. In contrast to these broadband designs, the present work focuses on realizing a narrowband resonance response using MXene by integrating the physics of bound states in the continuum (BIC). The integration of BIC-driven resonance mechanisms with MXene’s unique optoelectronic properties is anticipated to enable highly selective and efficient light–matter interactions, thereby paving the way for the development of next-generation flexible and high-performance sensing platforms.
\begin{figure*}[ht!]
 \centering
 \includegraphics[width=\textwidth]{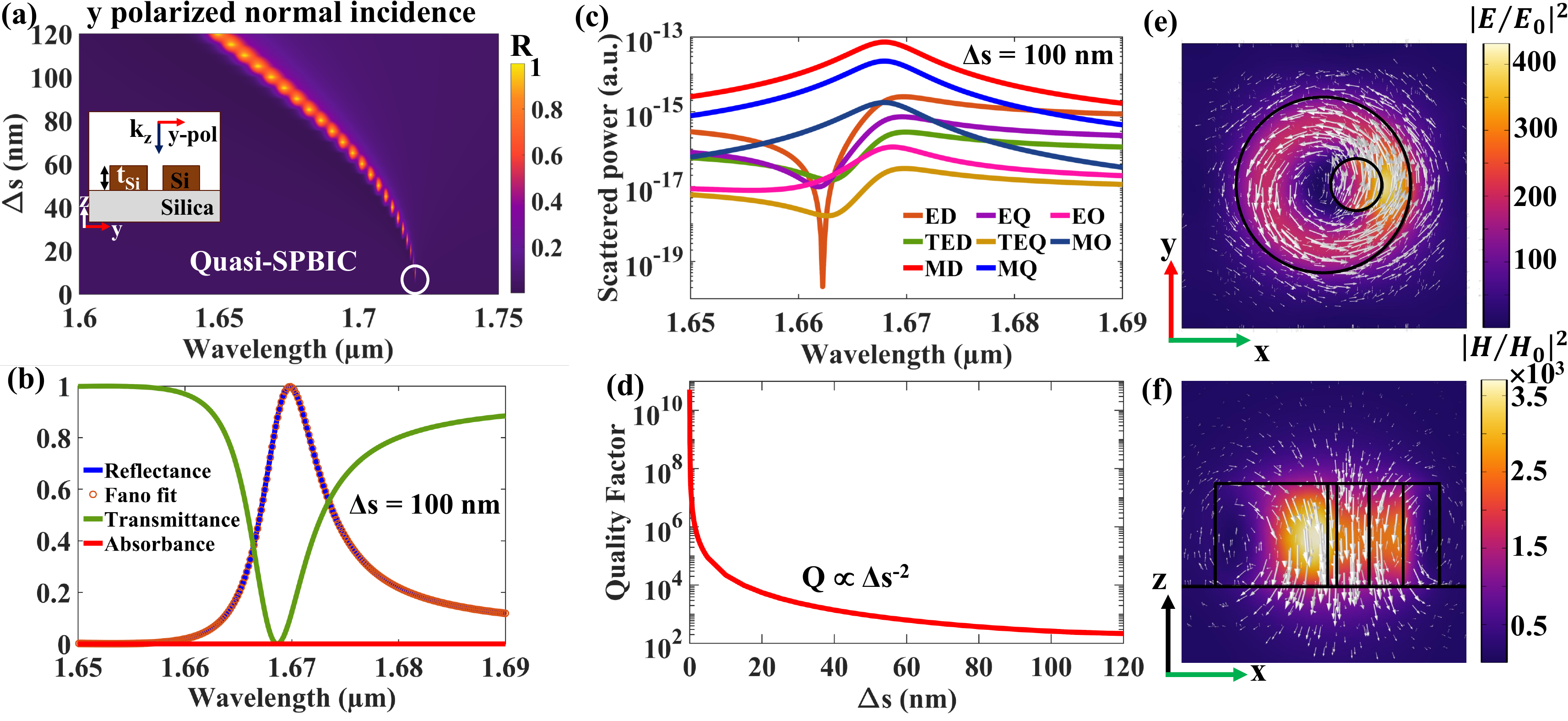}
 \caption{Spectral characteristics of Si nanodisk resonator with an offset air-channel (a) Reflectance plot for various offset ($\Delta s$) values of air-channel in the absence of MXene layer, constituting quasi-SPBIC for y polarized normal incident. The inset shows the cross-section of the geometry in the yz plane without MXene.(b) Spectral response with the Fano fit. (c) Multipolar component of the quasi-BICs resonance. (d) Quality factor for various $\Delta s$. (e) electric and (f) magnetic field intensity distribution along with the field vector in the near-field. For (b)-(f), graphs are for $\Delta s$ = 100 nm, except to (d).}
 \label{fig3}
\end{figure*}

 In this work, two design configurations are investigated in-depth to achieve high absorption over a narrow spectral range. In the first design, MXene is used as a thin film sandwiched between the dielectric spacer layer and substrate, whereas the second design has a composite nanodisk, made from three materials, namely Mxene, Silica spacer and Si.  Both designs exhibit ultra-high absorption, which is given by the physics of BIC combined with the plasmonic response of MXene. Both the design configurations are different with nearly similar geometrical parameters, yielding a similar narrowband optical response. These designs, when integrated with the Cryptophane-E, a compound known for its selectivity for alkane, can be used for Methane sensing with high sensitivity.

\section{Metasurface Designs and Working Principle}

Figure 1 shows the 3D schematic of both hybrid metasurface designs and the top view of the unit cell with geometrical parameters. The design 1 metasurface comprises silicon nanodisk resonators, where individual nanoresonators have a thorough air-hole channel offset from the nanodisk's center. 
 The dielectric spacer of a thickness ($t_{SiO_2}$) is sandwiched between the metasurface and Mxene thin film ($t_{mx}$). The whole structure is placed on a Silica substrate, as depicted in Figure 1a. In the second design configuration (Design 2), the composite nanodisk made up of MXene, a dielectric spacer, and Si (viewed from top to bottom) is placed over the silica substrate, as shown in Figure 1b. The composite nanodisk in design 2 also has an offset air channel.
 The design parameters have been optimized in such a way that both metasurface designs yield high absorption. Note that in the optimization, we have not employed any optimization algorithm, but rather used a trial-and-error approach. The geometrical parameters in both designs are periodicity (\textit{$\Lambda$}) = 865 nm,  Si nanodisk of radius ($r_{out}$) = 268 nm, and air hole radius ($r_{in}$) = 80 nm. The Si nanodisk has a thickness of ($t_{Si}$) = 242 nm. However, the thickness of the spacer is slightly different in both designs, $t_{SiO_2}$ = 200 nm in design 1 and $t_{SiO_2}$ = 320 nm in design 2.
 For designs 1 and 2, the amorphous Silicon is considered, whose refractive index is taken from Pierce and Spicer \cite{PhysRevB.5.3017}. The dielectric spacer and substrate have a constant refractive index of 1.44.  Here, Titanium Carbide MXene is chosen as the material for the plasmonic film. $\ce{Ti_3C_2}$ is one of the most investigated MXenes since its discovery because of its high conductivity and stability. The spectral dependence of the real and imaginary parts of the dielectric permittivity of the MXene is adopted from Karabchevsky's group paper \cite{https://doi.org/10.1002/adma.202210216}. At the interface, MXene thin film can exhibit surface plasmon polaritons (SPPs). However,  in a structured form of a nanoresonator, localized surface plasmon resonance (LSPR) in near-IR and mid-IR can be observed as the real part of the permittivity is negative. It is found that the wavelength at which the negative permittivity originates decreases with the increase in the thickness of the MXene film.  Therefore, thick MXene films exhibit significant metallic behavior \cite{doi:10.1021/acsphotonics.7b01439}. The optical properties of the MXene crucially depend on the thickness of the film. \\
 Figures 2a and c illustrate various intermediate steps to arrive at hybrid metasurface designs. For every intermediate design step, spectral responses are obtained and plotted in Figure. 2. 
 In design 1, a highly reflective nanostructure is combined with a plasmonic film to achieve maximum absorption, as shown in Figure 2b. Ideally, the addition of the perfect reflector blocks all the transmission and converts the two-port system into a one-port, maximizing absorption.
In the one-port system, the absorption of the single mode resonating at  $\omega_0$  can be obtained through coupled mode theory (CMT) \cite{doi:10.1021/ph400090p}:
\begin{equation}
    A=\frac{4\gamma_n\gamma_r}{(\omega-\omega_0)^2+(\gamma_n+\gamma_r)^2}
     \label{Eq1}
\end{equation}
where, $\omega$ is angular frequency of incident light, $\omega_0$ is angular resonant frequency, $\gamma_n$ is the dissipative loss rate, and $\gamma_r$ is radiative decay rate. 
The total quality factor ($Q_{total}$) of a particular mode can be calculated as \cite{https://doi.org/10.1002/adom.201900383}
\begin{equation}
    \frac{1}{Q_{total}}=\frac{1}{Q_{rad}}+\frac{1}{Q_{non-rad}}
    \label{Eq2}
\end{equation}
where $Q_{rad}$ is the radiative Q-factor in the absence of material and scattering losses,  $Q_{rad}$ = $\omega_0/2\gamma_r$. Note that scattering losses may arise from the fabrication imperfections, finite array size, and surface roughness. In simulations, scattering loss is neglected, while the material loss contributes to the non-radiative Q-factor, $Q_{non-rad}$ = $\omega_0/2\gamma_n$. 
The condition $\gamma_n$ = $\gamma_r$ implies that $Q_{rad}$ = $Q_{non-rad}$  at $\omega$ = $\omega_0$, yielding the critical coupling to attain perfect absorption (100$\%$). As illustrated in Figure 2b and Figure 2e, when the metasurface without MXene is illuminated normally, it functions solely as a perfect reflector. While designing this metasurface, the concept of symmetry-protected quasi-BICs has been used. Since tuning the asymmetry parameter of quasi SPBIC can achieve a high Q-factor, it can be utilized to create narrow-band absorbers. The MXene layer acts as a thin metallic film, and at normal incidence, it strongly reflects and weakly absorbs with negligible transmittance.  When the dielectric metasurface is combined with the MXene film separated by a spacer layer, the resonant cavity exhibits high absorptance.  
 Introducing the spacer layer ensures that the mode does not distort \cite{doi:10.1021/acsphotonics.2c00901}. 
But in design 2, the highly reflective nanostructure is combined with the MXene disk and the spacer disk to obtain the maximum absorption, as shown in Figure 2f. This limits the system to being a two-port system in contrast to design 1, which is a one-port system.
In the two-port system, the absorption of the single-mode resonating at  $\omega_0$  can also be obtained through CMT.
\begin{equation}
    A=\frac{2\gamma_n\gamma_r}{(\omega-\omega_0)^2+(\gamma_n+\gamma_r)^2}
     \label{Eq3}
\end{equation}
So, in such cases when $\gamma_n$ = $\gamma_r$ implies that $Q_{rad}$ = $Q_{non-rad}$  at $\omega$ = $\omega_0$ at critical coupling, maximum absorption is $50\%$. The MXene disk alone acts as a plasmonic resonator at normal incidence, it strongly absorbs approximately $50\%$ and weakly reflects with negligible transmittance as shown in Figure 2d.  When the dielectric metasurface is combined with the MXene disk separated by a spacer disk, the resonant cavity exhibits high absorbance that is achieved by tuning the asymmetry parameter.  
In simulations, the periodic boundary condition is used along the x and y axes. In contrast, perfectly matched layer boundary conditions are used along the z-axis of the unit cell. It must be noted that the normal incident and the y-polarized light have been considered in both cases. 

\section{Results and Discussion}

\subsection{Reflective Metasurfaces without MXene}
The proposed design of the nanoresonator (air-hole in nanodisk) possesses $C_{2v}$ symmetry in the xy plane when the air channel is not displaced from the center and there is no substrate. Such a structure supports an ideal-SPBIC. In the presence of a substrate, the symmetry in the xy plane is broken, turning the SPBIC into a quasi-SPBIC. Interestingly, the symmetry in the $x$ = 0 plane is also broken when the air channel is displaced from the center by $\Delta s$ such that the symmetry $C_{2v}$ converts to $C_s$. Due to this symmetry breaking, the spectral resonance feature that was not seen in the spectral response becomes visible prominently as the Fano resonance. For the optimized parameters mentioned above, the structure exhibits only one mode, a symmetry-protected quasi-BIC named quasi-SPBIC.

In case of zero offset of air-channel (i.e. $\Delta s$ = 0) in the metasurface, no resonance occurs for y-polarized normal incidence, indicating complete decoupling of the mode with the incidence light (marked as the white circle in Figure 3a). As soon as the air-hole position in the nanoresonator is shifted, i.e., an increase in $\Delta s$, the in-plane symmetry is broken, and the resonance appears in the reflectance spectrum, as seen from Figure 3a, which is mainly due to enhanced coupling between the incidence light and the radiating mode. This also decreases the linewidth of the resonance. Figure 3b shows the complete spectrum, i.e., reflectance, transmittance, and absorbance at $\Delta s$ = 100 nm. We can observe how a lossless Si metasurface acts as a perfect reflector at $\lambda$ = 1.67 $\mu$m with no absorption. The lossless Si contributes to the radiative Q-factor ($Q_{rad}$) as there is no material loss in Si in the near-IR. Hence, the total Q-factor ($Q_{total}$) is driven by $Q_{rad}$ such that with an increase in structural perturbation ($\Delta s$), the Q-factor decreases. When the $\Delta s$ = 100 nm, the metasurface has a Q-factor of $\approx$ 261. The Q-factor exhibit the inverse quadratic relation with the asymmetry, $\Delta s$ as $Q$ = $k/\alpha^2$, where k is the proportionality constant dependent on metasurface design and $\alpha$ is asymmetry parameter given as $\alpha = \Delta s/r_{out}$ or $Q$ = $kr_{out}^2/\Delta s^2$  as shown in Figure 3d \cite{PhysRevLett.121.193903}. 
The near-field distribution of electric, magnetic fields and the Cartesian multipole decomposition unveil the quasi-SPBIC resonance characteristics. The polarization induced by the incident light has also been incorporated in calculating the scattered power all multipole moments of a meta-atom (nanoresonator) unit cell.  The scattered power of each multipole can be calculated from equations provided in the supplementary material Section S1. Here, ED is an electric dipole, TED is a toroidal electric dipole, MD is a magnetic dipole, EQ is an electric quadrupole, TEQ is a toroidal electric quadrupole, MQ is a magnetic quadrupole, EO is an electric octupole, and MO is a magnetic
octupole, respectively.
Each mode of a nanoresonator is composed of several multipoles and contributes to far-field radiation. However, in the case of symmetry-protected BICs, the dominant component of the multipole moment is along the direction of incident light, preventing them from coupling with each other. In this particular study, the magnetic dipole moment of the mode aligns to the z-axis ($\ce{MD_z}$) as its dominant component does not radiate in the far field. In contrast, the effect of other moments is minimal \cite{PhysRevB.100.115303}. As a result, the nanoresonator demonstrates the formation of an ideal BIC. When the symmetry is perturbed, for example, at $\Delta s$ = 100 nm, other multipole moments also become strong, and the mode becomes leaky in the far field. It is observed from Figure 3c that the MD and MQ are dominant, whereas other multipoles also exhibit peaks at the resonant wavelength. Thus, the SPBIC transforms into a radiating quasi-SPBIC with high reflectance. The near-field distributions of the electric and magnetic fields shown in Figure 3e and Figure 3d also confirm MD as a dominant mode, where the circulating electric field gives rise to the out-of-plane magnetic field in the xy plane. 

\subsection{Absorptive Metasurface with MXene}

\begin{figure*}[ht!]
 \centering
 \includegraphics[width=\textwidth]{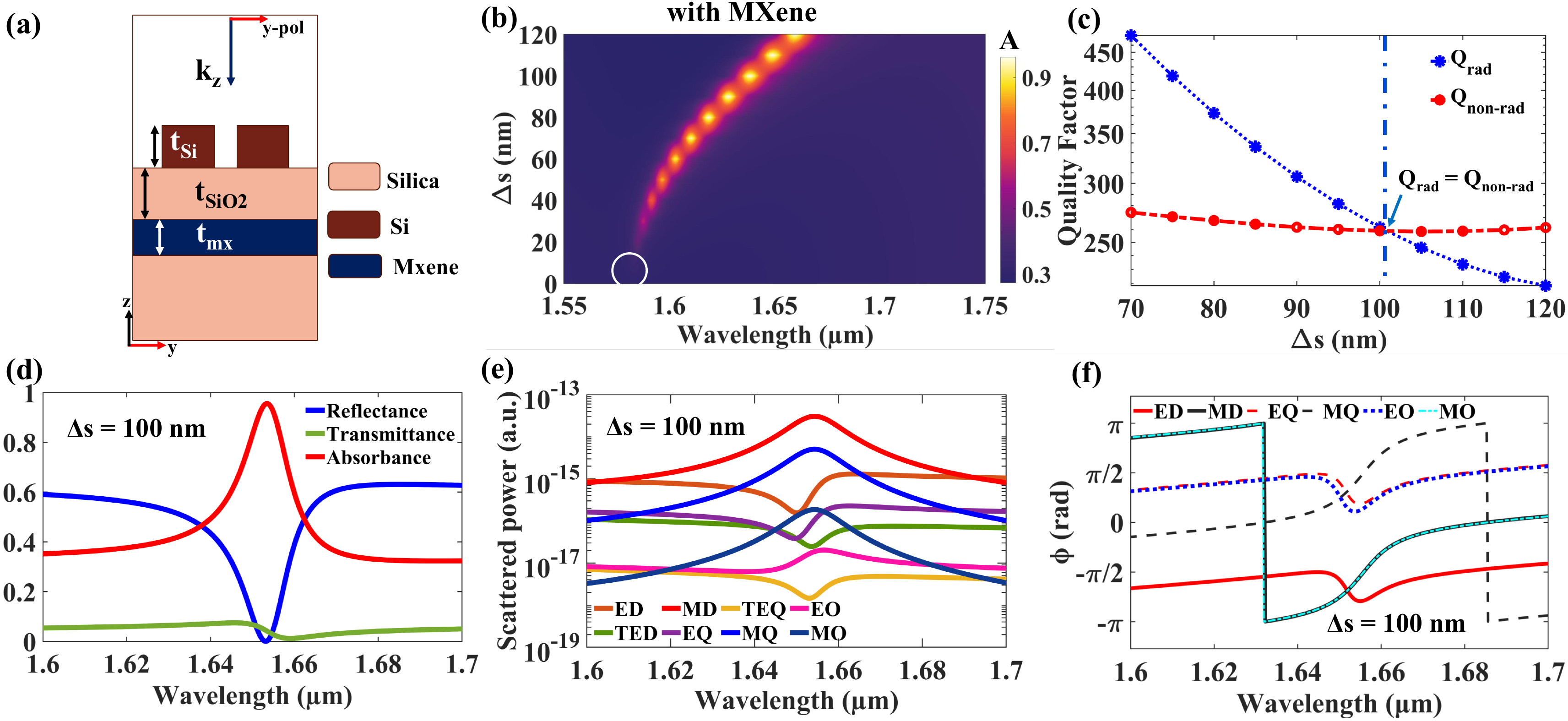}
 \caption{(a) Cross-section of the geometry in the yz plane with MXene layer as an absorptive thin film, (b) Absorptance surface plot, (c) Radiative and non-radiative quality factor for various $\Delta s$ values. (d) Spectral (transmittance, reflectance, and absorptance) characteristics at $\Delta s$ = 100 nm. (e) multipole contributions when the metasurface is kept in the air for $\Delta s$ = 100 nm. (f) phase plots of various multipole (ED, MD, EQ, MQ, EO, and MO) moments.}
 \label{fig4}
\end{figure*}
\begin{figure*}[ht!]
 \centering
 \includegraphics[width=\textwidth]{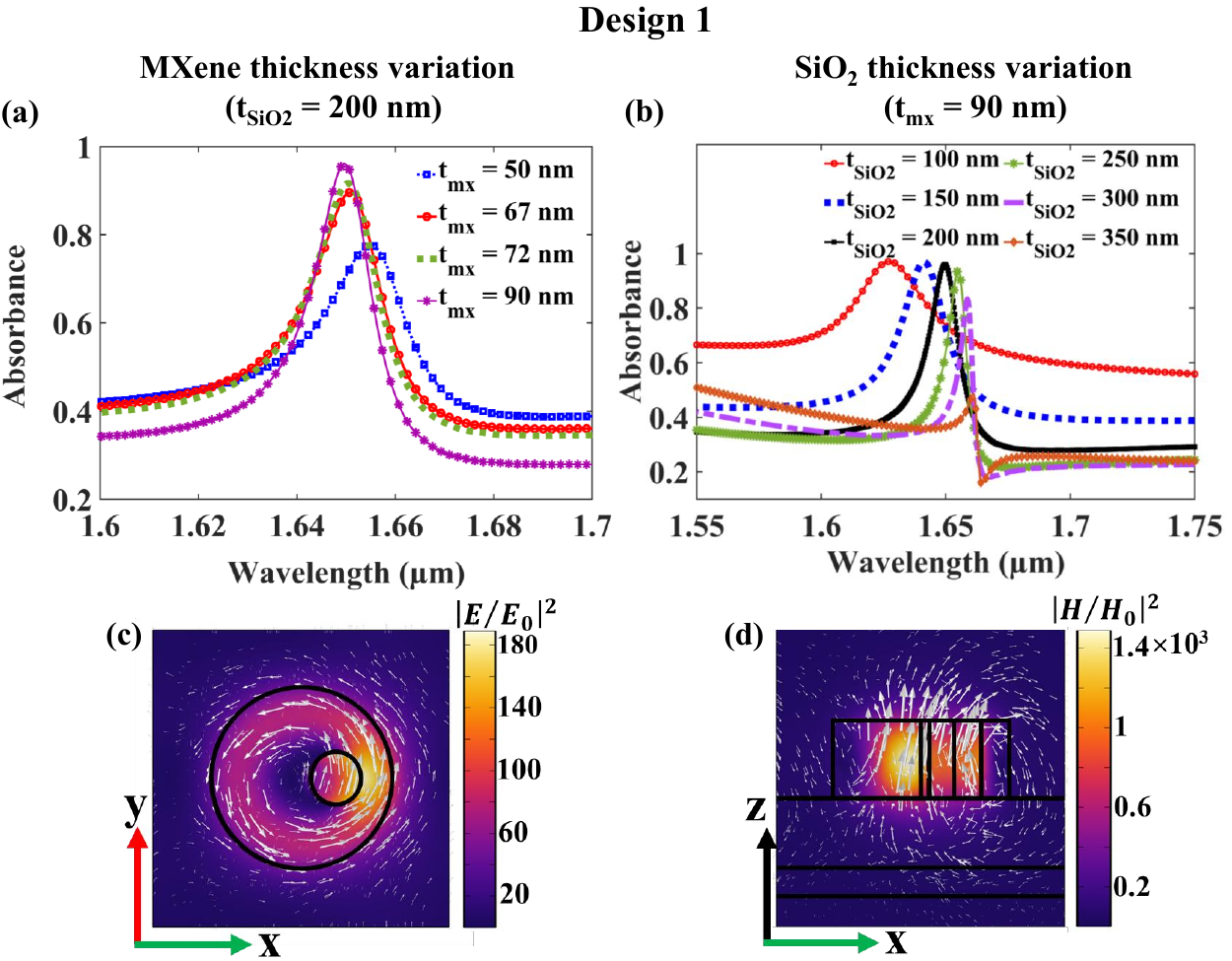}
 \caption{(a) Effect of MXene layer thickness when spacer layer is fixed to ${t_{SiO_2}}$ = 200 nm. (b) Impact of spacer layer thickness for a fixed MXene layer of ${t_{mx}}$ = 90 nm. (c) Electric and (d) magnetic field distributions along with vector plot at $\lambda$ = 1653 nm for ${t_{mx}}$ = 90 nm and ${t_{SiO_2}}$ = 200 nm.}
 \label{fig5}
\end{figure*}
The absorption in the Si nanodisk can be enhanced by combining it with the lossy material. In order to do so, we have studied two design aspects as depicted in Figure 1.
\subsubsection{Absorption in Design 1}
The absorption in the reflective metasurface of design 1, as discussed earlier, can be enhanced by combining the dielectric metasurface with a plasmonic layer, which is 90 nm thick MXene layer and a spacer layer $\ce{SiO_2}$ of thickness, $t_{SiO_2}$= 200 nm, as shown in the schematic cross-section of the design in Figure 4a.  

The light absorption in the dielectric metasurface is mainly controlled by the three parameters (i) asymmetry parameter, $\Delta s$, (ii) thickness of spacer layer, $t_{SiO_2}$, and (iii) thickness of MXene layer, $t_{mx}$. 
Due to plasmonic behavior in the wavelength of interest, MXene introduces a necessary material loss for the metasurface to achieve absorption and prevent light transmission. For y-polarized normal incidence, the silicon nanoresonator (nanodisk with air channel) with in-plane symmetry, along with MXene, does not exhibit absorption (Figure 4b) as the MD mode fails to couple with the incident light. Additionally, the radiative Q-factor ($Q_{rad}$) is significantly higher than the non-radiative Q-factor ($Q_{non-rad}$). As the value of $\Delta s$ increases, the absorption increases, reaching maximum absorption around $\Delta s$ = 100 nm.  Subsequently, absorption decreases, as depicted in Figure 4b. 

We use eigenmode analysis to calculate the radiative Q-factor ($Q_{rad}$) by assuming the imaginary part of the refractive index of the MXene layer is zero for various values of $\Delta s$. Subsequently, the total Q-factor ($Q_{total}$) is computed, considering the loss in the MXene layer. The non-radiative Q-factor ($Q_{non-rad}$) is derived using the relationship specified in Eq. 2. From Figure 4c, it is observed that $Q_{rad}$ decreases as $\Delta s$ increases, whereas $Q_{non-rad}$ remains almost constant, as it mainly depends on the material loss and the thickness of the MXene layer. At $\Delta s$ = 100 nm, both the $Q_{rad}$ $\approx$ $Q_{non-rad}$ indicate the critical coupling such that light does not scatter into the surrounding medium and almost perfect absorptance is achieved.

The hybrid metasurface (design 1) exhibits an absorptance peak of 95.6$\%$, transmittance of 0.04$\%$, and minimal reflectance when $\Delta s$ = 100 nm, as seen clearly in Figure 4d. The metasurface yields a total Q-factor of 129. The MXene film not only behaves as a perfect reflector, but also absorbs and transmits a certain fraction of light, as depicted in Figure 2c. This characteristic persists during the interaction of light with the MXene film, the Si nanodisk, and the dielectric spacer. 

The multipole decomposition technique helps to analyze and comprehend the role of multipole moments in absorption.  In the analysis, it was observed that magnetic multipoles (MD, MQ, and MO) exhibit a peak at a wavelength of 1653 nm (Figure 4e), even in the absence of MXene. On the other hand, the electric multipole moments (ED, EQ, except EO) display a dip in the scattered power plot, indicating reflectance suppression. This observation becomes more evident when examining the phase spectrum of the multipole moment (Figure 4f), where the magnetic and electric multipoles exhibit opposite slopes, suggesting the destructive interference that suppresses backward scattering (i.e., reflectance).
The suppression of reflectance and blocking of transmittance leads to the high absorptance at $\lambda$ = 1653 nm. \\
The plasmonic behavior of MXene can be significantly improved by increasing the thickness of the MXene film. As shown in Figure  5a, when the MXene thickness is increased from 50 nm to 90 nm while $\ce{SiO_{2}}$ thickness is fixed at 200 nm, the resonant wavelength blue-shifts, and the absorbance approaches unity while the transmittance nearly approaches zero. As the MXene film thickness increases, the transmission becomes minimal, potentially resulting in perfect absorbance in the metasurface. Since the interlayer coupling between the MXene layers is enhanced, the conductivity of the MXene film is also boosted. 
The addition of a spacer layer significantly enhances the absorption by creating a cavity between the dielectric metasurface and the MXene film, such that the round-trip phase shift leads to destructive interference and suppresses the scattering in the backward direction. By varying the thickness of $\ce{SiO_{2}}$, the optimum value corresponding to nearly perfect absorption is determined, as shown in Figure 5b. At $t_{SiO_2}$ = 200 nm, design 1 exhibits maximum absorptance with a total Q-factor of 129. Conversely, broadband behavior is observed at $t_{SiO_2}$ = 100 nm. The absorption reduces when the thickness decreases from 200 nm. The near-field distribution of electric and magnetic fields at the absorption wavelength is illustrated in Figure 5c and Figure 5d, respectively. It is observed that near-field electric field lines maintain a configuration similar to that of the case discussed earlier. However, the magnetic field lines in the xz plane are slightly tilted away from the z-axis, indicating energy confinement in the spacer region.

\begin{figure*}[ht!]
 \centering
 \includegraphics[width=\textwidth]{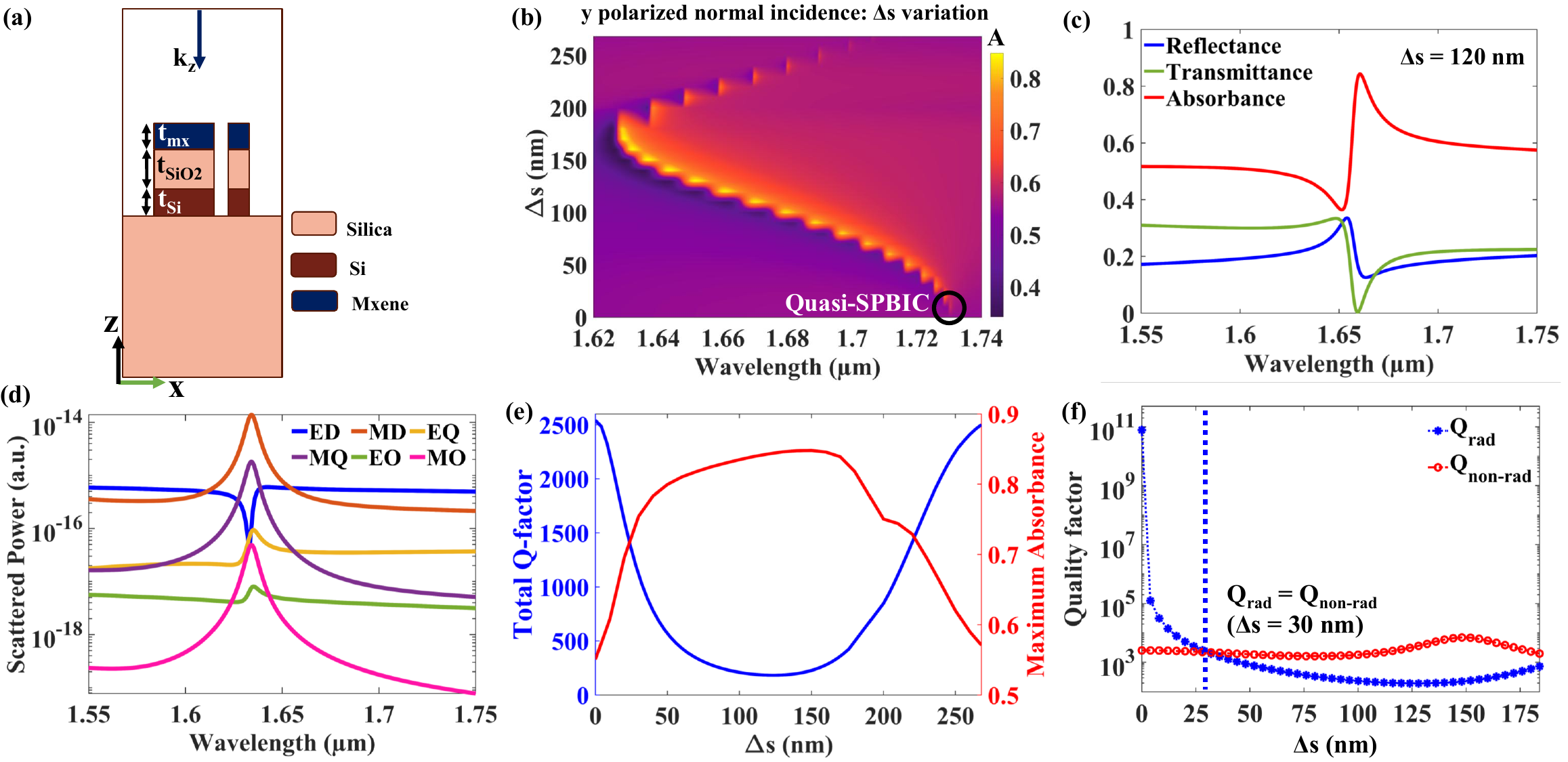}
 \caption{(a) Cross-section of the geometry in the xz plane with MXene on top of nanodisk (design 2), (b) Absorbance plot for various values of air-channel shift, (c) Spectral (transmittance, reflectance, and absorptance) characteristics at $\Delta s$ = 120 nm, (d) multipole contributions in absence of substrate at $\Delta s$ = 120 nm. (e) total quality factor and maximum absorbance, and (f) Radiative and non-radiative quality factor for various $\Delta s$ values.  }
 \label{fig6}
\end{figure*}
\begin{figure*}[ht!]
 \centering
 \includegraphics[width=\textwidth]{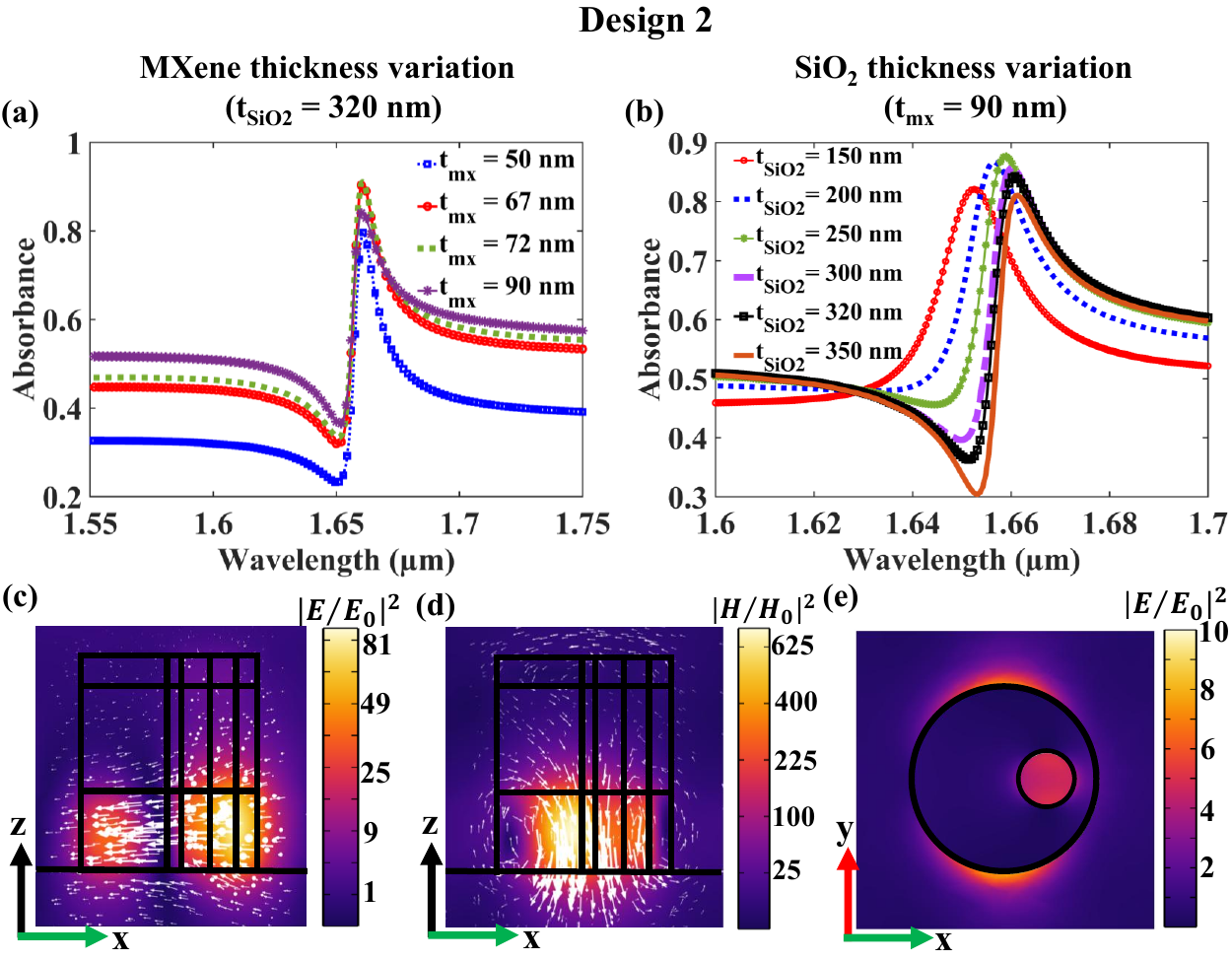}
 \caption{Parameteric study of design 2 hybrid metasurface, (a) Effect of MXene layer thickness when the spacer layer thickness is fixed to 320 nm, (b) Effect of spacer layer thickness for 90 nm thick MXene layer, (c) Electric field, (d) magnetic field distributions and the vector plot in xz-plane and (e) electric field distribution in MXene aided nanodisk in xy-plane at $\lambda$ = 1653 nm when ${t_{mx}}$ = 90 nm and ${t_{SiO_2}}$ = 320 nm.}
 \label{fig7}
\end{figure*}

\subsubsection{Absorption characteristics of Design 2 metasurface}
So far, we have established that the absorption in design 1 metasurface is mainly driven by the structural asymmetry that gives rise to quasi-BIC, the losses associated with MXene, and the spacer layer. In design 2, MXene nanodisk is combined with $\ce{SiO_2}$ and Si nanodisk, which gives rise to high absorption. MXene being a plasmonic material in NIR may exhibit LSPR in NIR when nanostructuring is done \cite{chaudhuri2018highly}. In contrary, in studied design 2, the MXene is so thin (= 90 nm) that it gives rise to the broadband absorption of $\sim50\%$ in the wavelength range of interest. 

As established earlier (see Figure 3), Si nanodisk exhibits reflectance when there is no material loss involved under y-polarized normal incidence. When MXene is coated on silica-silicon nanodisk with a presence of an air-hole at the center (see Figure 6a), the metasurface doesnot exhibit narrow absorption resonance. However, as the air channel shifts by $\Delta s$ in the x direction, the BIC mode converts to QBIC with a Fano lineshape, as observed from Figure 6b. The absorption increases with the increase in $\Delta s$, and becomes maximum ($\approx 85\%$) at $\Delta s$ = 120 nm at $\lambda$ = 1.66$\mu$m. The spectral resonance at this wavelength shows a Q-factor 184. Any further rise in $\Delta s$ decreases the absorption as observed from Figure 6b. It is also observed from Figure 6c that the difference between the peak and the dip of the Fano resonances in absorption is $\sim 40\%$ at $\lambda$ = 1.655 $\mu$m with negligible transmission and $\sim15\%$ reflection whereas in the vicinity of the resonance, the design 2 exhibits a constant absorption of nearly $\sim50\%$, which is in contrast with design 1. The dominant multipole in design 2 is also magnetic dipole which governs the resonance as shown in Figure 6d.
Through eigemode analysis and Fano fitting, the total Q-factor is calculated and shown in Figure 6e along with the maximum absorbance with respect to air-channel shift. It is observed that the maximum absorption occurs when the shift is nearly $\Delta s$ = 120nm, but has a low Q-factor. To attain the high-Q resonance, absorption should be compromised. The increase in absorption with an increase in the asymmetry parameter implies that the strong coupling occurs between the non-radiative losses of the MXene and the radiative losses of the quasi-BIC resonance.  

Since MXene has been used in the form of a nanodisk rather than a thin film (design 1), the whole configuration works as a two-port system according to the Eq. 3. From the radiative and non-radiative Q-factor plot (Figure6f),  the two-port system exhibits the $\approx 50\%$ absorption when $Q_{rad} = Q_{non-rad}$, which occurs at $\Delta s$ = 30 nm, implying the critically coupled state. 

When the MXene thickness is changed (from 50 nm to 72 nm) while fixing all other parameters, it is found that absorption increases for a fixed spacer thickness of $t_{SiO_2}(2)$ = 320 nm (Figure 7a), but decreases after 72 nm without any change in resonance wavelength. The maximum absorption of $\sim90\%$ occurs at $t_{mx}$ = 72 nm. To further enhance the absorption in such a hybrid design, structural parameters can be fine-tuned or better optimized through an optimization algorithm. It is noteworthy to state that the spacer plays a very important role in achieving high absorption, more specifically for design 2 hybrid metasurface, as illustrated in Figure 7b, where the absorption is significantly enhanced due to the momentum matching between the plasmon of the MXene disk and the photonic modes in the Si disk. This fact can further be validated by the electric field distribution (Figure 7c), which mainly confines at the periphery of the Si nanodisk and circulates clockwise in the x-y plane. The circulating electric field produces an out-of-plane magnetic field, as shown in Figure 7d. Figure 7e, confirms the presence of the electric field at the edges of the MXene disk and in the air channel, affirming the formation of LSPR. Also, note that the electric field in the $\ce{SiO_2}$ disk primarily concentrates the air channel, facilitating the coupling between the modes of MXene and the Si nanodisk. 

\subsection{Application of Designed Metasurface as Methane Gas Sensor}

\begin{figure*}
 \centering
 \includegraphics[width=\textwidth]{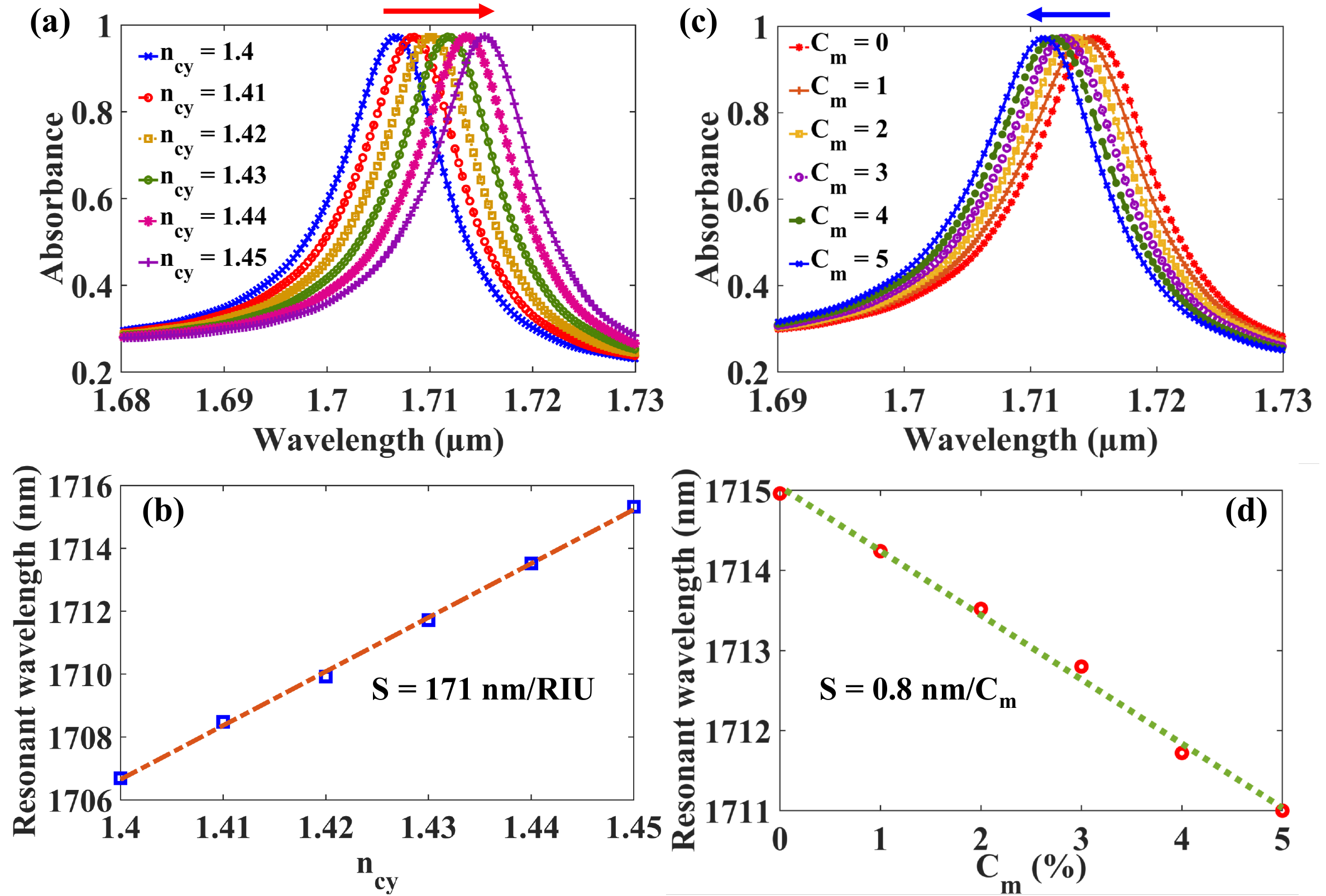}
 \caption{(a) Absorbance plot dictating the refractive index sensing for a 400 nm thick analyte layer, (b) Sensitivity plot for bulk refractive index sensing, (c) Absorbance plot when a 400 nm thick Cryptophane-E doped polymer layer is exposed to different concentrations of Methane gas. The blue arrow represents the blue shift in the spectral response. (d) Methane gas sensitivity plot for various concentrations of the gas.}
 \label{fig8}
\end{figure*}
Refractive index-based sensing has been widely utilized in various applications to detect changes subject to variation in the refractive index of the surrounding medium. This method relies on the interaction between light and matter; specifically,  the sensitivity of the evanescent waves around the metasurface is sensitive to the change in the refractive index of the surrounding media \cite{SHARMA2020100044}.  The primary challenge in designing a highly sensitive device is effectively responding to resonances influenced by near-field coupling effects. 
The sensors' performance is assessed using certain criteria such as selectivity, sensitivity ($S$), and figure of merit (FOM). Both sensitivity and FOM are defined as \cite{SHARMA2020100044,ben2022high}:
\begin{equation}
   S = \Delta \lambda/\Delta n
   \label{eq4}
   \end{equation}
   \begin{equation}
   FOM = \frac{S}{FWHM}
   \label{eq5}
\end{equation}
where $\Delta \lambda$ is the shift in wavelength, corresponding to a change in refractive index, $\Delta n$, and FWHM is the full-width half-maximum of the spectral resonance.
We investigate the hybrid metasurface design 1 where MXene behaves as a thin film and yields bulk refractive index sensing when the surrounding refractive index ($n_{cy}$) is varied from 1.4 to 1.45, as shown in Figure 8a. Numerical simulations dictate unity absorption of the incident electromagnetic field. In simulations, a polymeric film is considered as an analyte layer whose thickness is 400 nm. Figure 8a shows a clear redshift in the spectral response. The metasurface design 1 demonstrates a sensitivity of $S =  $ 179 nm/RIU and FOM = 17.56 $\ce{RIU^{-1}}$ for $\Delta n =0.01$. In Figure 8b, it is evident that the average sensitivity is 171 nm/RIU. The sensitivity enhancement is attributed to the asymmetric excitation of both electric and magnetic resonances. \\
In the near-infrared region, Methane gas exhibits two near-infrared vibrational absorption bands, $\nu_{2}$+2$\nu_{3}$ and 2$\nu_{3}$, with corresponding absorption peaks in the wavelength range of 1330-1335 nm and 1650-1670 nm \cite{hong2020state}. The Methane gas absorption band at 1653 nm (2$\nu_{3}$) is significantly stronger by an order of magnitude and experiences minimal interference from other gases compared to the weaker ($\nu_{2}$+2$\nu_{3}$) band. This makes it highly suitable for direct spectroscopic sensing applications \cite{wang2013fiber}.\\
The effective method to enhance the selectivity and sensitivity of gases from the mixtures is to use a gas-sensitive material on top of the metasurface. In the case of Methane gas sensing, a gas-permeable polymer, such as polysiloxane doped with the Cryptophane-E, is one of the primary molecules that facilitates the interaction with Methane gas\cite{Yang:10}. Notably, Cryptophanes are the sole functional materials that exhibit a photosensitive response to Methane \cite{benounis2005study}. When Methane gas is adsorbed by the Cryptophane-doped polymer layer, gas enters the inner cavity of the Cryptophane molecule, causing a change in the molecule's dipole moment, leading to a corresponding alteration in the refractive index. 
The refractive index $n$ of Cryptophane-E doped polymer thin film variation with Methane concentration  $C_m$, is given as \cite{zhang2015measurement}
\begin{equation}
    n = 1.448 - 0.46C_m
    \label{eq6}
\end{equation}

Figure 8c illustrates the absorbance spectra of the hybrid metasurface design 1 for various concentration values of Methane gas. 
When the Methane gas concentration $C_m$ increases, the resonance characteristic remains consistent, but the resonant wavelength experiences a blue shift. At very low Methane concentration ($C_{m}$ = 0), the Cryptophane-doped polymeric film behaves similarly to an analyte with a refractive index of 1.45. As the Methane concentration increases by 1$\%$, the resonance shifts to a shorter wavelength due to the reduced refractive index of the Cryptophane-E doped polymer layer. In this case, the expression for the sensitivity is modified to 
\begin{equation} S = \frac{\Delta \lambda}{ \Delta C_{m}} = \frac{\Delta \lambda}{\Delta n}\frac{\Delta n}{\Delta C_{m}} 
\label{eq7}
\end{equation} 
where $\Delta C_{m}$ denotes the change in the Methane concentration. 
When the Methane gas concentration changes by 1$\%$, the resonant wavelength is blue-shifted by 0.72 nm. The average sensitivity is 0.8 nm per 1$\%$ change in Methane gas concentration with FOM = 0.07 per 1$\%$ change in Methane concentration, as depicted in Figure 8d. Simulation results suggest an enhancement in Methane gas sensitivity when cryptophane E is used over the hybrid metasurface. 

\section*{Conclusions}
In this work, we demonstrated two novel MXene–dielectric hybrid metasurface absorbers that utilize symmetry-protected bound states in the continuum physics to achieve narrowband and high-Q resonances suitable for Methane detection in the near-infrared region. By carefully engineering the structural asymmetry and integrating MXene either as a thin plasmonic layer (Design 1) or as a nanodisk resonator (Design 2), we achieved strong light–matter interactions with near-unity absorption. Design 1 reached an absorbance of $>90\%$ with a Q-factor of 129. In contrast, Design 2 attained $>80\%$ absorbance with a Q-factor of 184, confirming that the interplay between radiative and non-radiative loss channels can be finely tuned for optimal performance.

The sensing functionality was further enhanced by incorporating a Cryptophane-E layer, whose high affinity for Methane enables selective adsorption and modulation of the index. Numerical simulations revealed a bulk refractive index sensitivity of $\sim171$ nm/RIU and FOM = 17.56 \ce{RIU^{-1}}and a Methane-specific sensitivity of 0.8 nm per unit concentration percentage, establishing the hybrid metasurface as a competitive platform for trace-level gas monitoring. The device operates around 1653 nm, precisely overlapping with the strong overtone absorption band of Methane, thereby maximizing selectivity and minimizing cross-sensitivity to other gases. Note that the hybrid metasurface design concepts can also be extended to other greenhouse gases for environmental monitoring, molecular spectroscopy, and photonic on-chip integration.



\section*{Author contributions}
Authors S.S. performed designing, simulation, and analysis, and M.P. assisted in the analysis. S.S. and M.P. carried out the manuscript writing with input from co-authors. Av. K helped in technical discussions and provided the data of MXene; A.K. and S.K.V. supervised the research, monitored the overall progress of the work, and manuscript writing. 

\section*{Conflicts of interest}
There are no conflicts to declare.

\section*{Data availability}
The data will be made available upon reasonable request.


\medskip
\textbf{Supporting Information} \par 

\section*{Acknowledgements}
This research was supported by the Ministry of Science $\&$ Technology, Israel-India Grant $\#0006276$ and DST Govt. of India, vide no: DST/INT/ISR/P-34/2023.  

\bibliographystyle{MSP}



\end{document}